\documentclass[prl,twocolumn]{revtex4}
\usepackage{epsfig}
\usepackage{bm}

\begin{document}

\title{Cluster-scaling, chaotic order and coherence in DNA}

\author{A. Bershadskii}

\affiliation{
ICAR, P.O. Box 31155, Jerusalem 91000, Israel
}

\begin{abstract}
Different numerical mappings of the DNA sequences have been studied using a new cluster-scaling method 
and the well known spectral methods. It is shown, in particular, that the nucleotide sequences in DNA 
molecules have robust cluster-scaling properties. These properties are relevant to both types of nucleotide 
pair-bases interactions: hydrogen bonds and stacking interactions. It is shown that taking into account the cluster-scaling properties can help to improve heterogeneous models of the DNA dynamics. It is also shown that a chaotic (deterministic) order, rather than a stochastic randomness, controls the energy minima positions of the stacking interactions in the DNA sequences on large scales. The chaotic order results in a large-scale chaotic coherence between 
the two complimentary DNA-duplex's sequences. A competition between this broad-band chaotic coherence 
and the resonance coherence produced by genetic code has been briefly discussed. The Arabidopsis plant genome 
(which is a model plant for genome analysis) and two human genes: BRCA2 and NRXN1, have been considered as 
examples.      
\end{abstract}

\maketitle

\section{Introduction}

A DNA molecule carries information in the form of four chemical groups or nucleotide bases: 
adenine, cytosine, guanine, and thymine, represented by the letters A, C, G and T. The order 
of bases on a DNA strand is the DNA sequence. If we read along one of the two DNA-helix sides we get 
text like GATACA... In the double-stranded DNA, the two strands run in opposite directions 
and the bases pair up such that A always pairs with T and G always pairs with C. That is because 
these particular pairs fit exactly to form effective hydrogen bonds with each other. 
The A-T base-pair has 2 hydrogen bonds and the G-C base-pair has 3 hydrogen bonds. 
The G-C interaction is therefore stronger than A-T, 
and A-T rich regions of DNA are more prone to thermal fluctuations and to 
initiation sites (origin) at unwinding stage of DNA replication process. 
The bases are oriented perpendicular to the DNA-helix axis. 
Constant thermal fluctuations result in local twisting, stretching, 
bending, and unwinding of the double-strands. 
\begin{figure} \vspace{-0.5cm}\centering
\epsfig{width=.45\textwidth,file=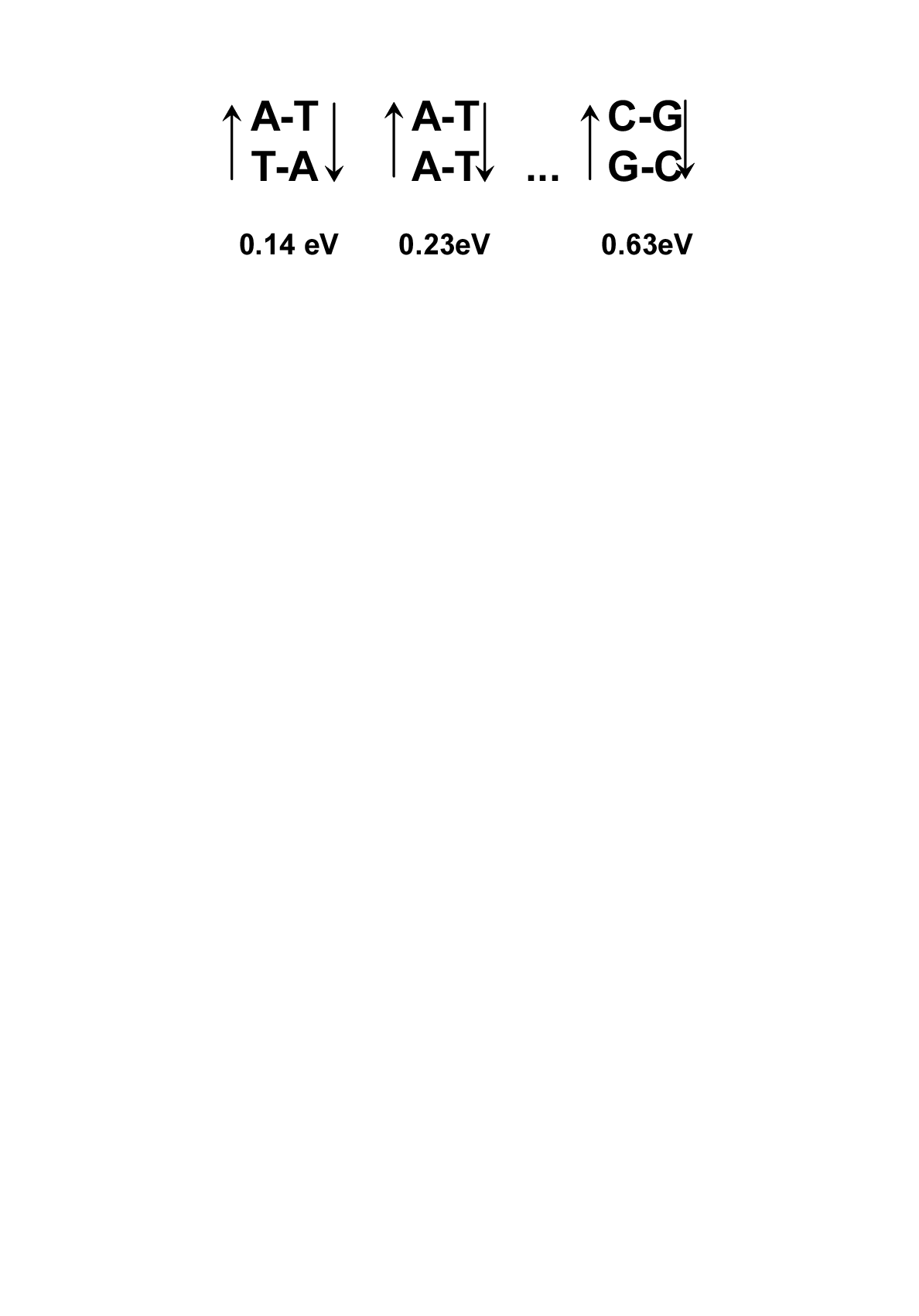} \vspace{-8.5cm}
\caption{The stacking energies for different stacked
base pairs.}
\end{figure}

In solution DNA assumes linear configuration because it is the one of minimum energy. 
The helix axis of DNA in vivo is usually strongly curved because the stretched length 
of the human genome, for instance, is about 1 meter and this length needs to be "packaged" 
in order to fit in the nucleus of a cell (the diameter of the nucleus from a typical human somatic cell 
is about $5 \times10^{-6}$ meters). Therefore, the DNA has to be highly organized. 
This packaging of DNA deforms it physically, thereby increasing its energy (less stable than relaxed DNA, 
due to less than optimal base stacking). In this situation certain strain is relieved by supercoiling: 
helix bends and twists to achieve better base stacking orientation despite having too many bp/turn. 
The difference in A-T and G-C interactions can be used for optimizing the free energy. 
The base-pairs stacking energies (the main stabilizing factor in the DNA duplex, see for instance 
Ref. \cite{yak}) are highly dependent on the base sequence \cite{sae}. These interactions come partly from the overlap 
of the $\pi$ electrons of the bases and partly from hydrophobic interactions. 
Quantum chemistry calculations give rather different
energies for different stacked base pairs: Fig. 1 (cf. Ref. \cite{p}). Therefore, certain clustering of the 
base-pairs can be used by nature in order to minimize the excess energy that builds up when 
DNA molecules are deformed during the process of packaging. The physical constraints given by the supercoiling of the DNA sequence, in particular to the positioning of nucleosomes along the sequence \cite{va}, play significant role in creating of the clustering.

Moreover, the increase in stored (potential) energy within  the molecule is then available 
to drive reactions such as the unwinding events that occur during DNA 
replication. Before replication of DNA can occur, the length of the DNA double helix about to be copied must 
be unwound and the two strands must be separated by breaking the hydrogen bonds that link the paired bases. 
The process of replication begins in the DNA molecules at thousands of sites called origins of replication. 
Because the location and time of initiation of origins is generally stochastic, the
time to finish replication will also be a stochastic process. The random distribution of origin firing raises the random gap problem: a random distribution will lead to occasional large gaps that should take 
a long time to replicate. Despite this each cell in a population must complete the
replication process in an accurate and timely manner (see for instance, Refs. \cite{hyr},\cite{jun}). 
Different solutions to this problem have been suggested (see, for instance, Refs. \cite{blow},\cite{rhi},\cite{con},\cite{yang}) . 

If the spacing of origins is not completely random then any regularity in the spacing 
of origins will tend to suppress the large gaps \cite{blow}. For instance, origins within specific 
clusters could be preferred to fire \cite{mes},\cite{she}. Since 
a G-C base pair, with three hydrogen bonds, is expected to be harder to break than an A-T base pair with only 
two bonds, a clustering of these two kinds of the base-pares can be operational in order to solve the random 
origin firing problem. The stacking interactions can also contribute to solution of this problem. 
It will be shown below that a chaotic (deterministic) order, rather than a stochastic randomness, 
controls the energy minima positions of the stacking interactions in the DNA sequences on large-scales. 
This chaotic order not only introduces a regularity into the spacing of the origins but also results in 
a long-range coherence between the two complimentary DNA-duplex's strands.

\section{Cluster-scaling}
\begin{figure} \vspace{-1cm}\centering
\epsfig{width=.45\textwidth,file=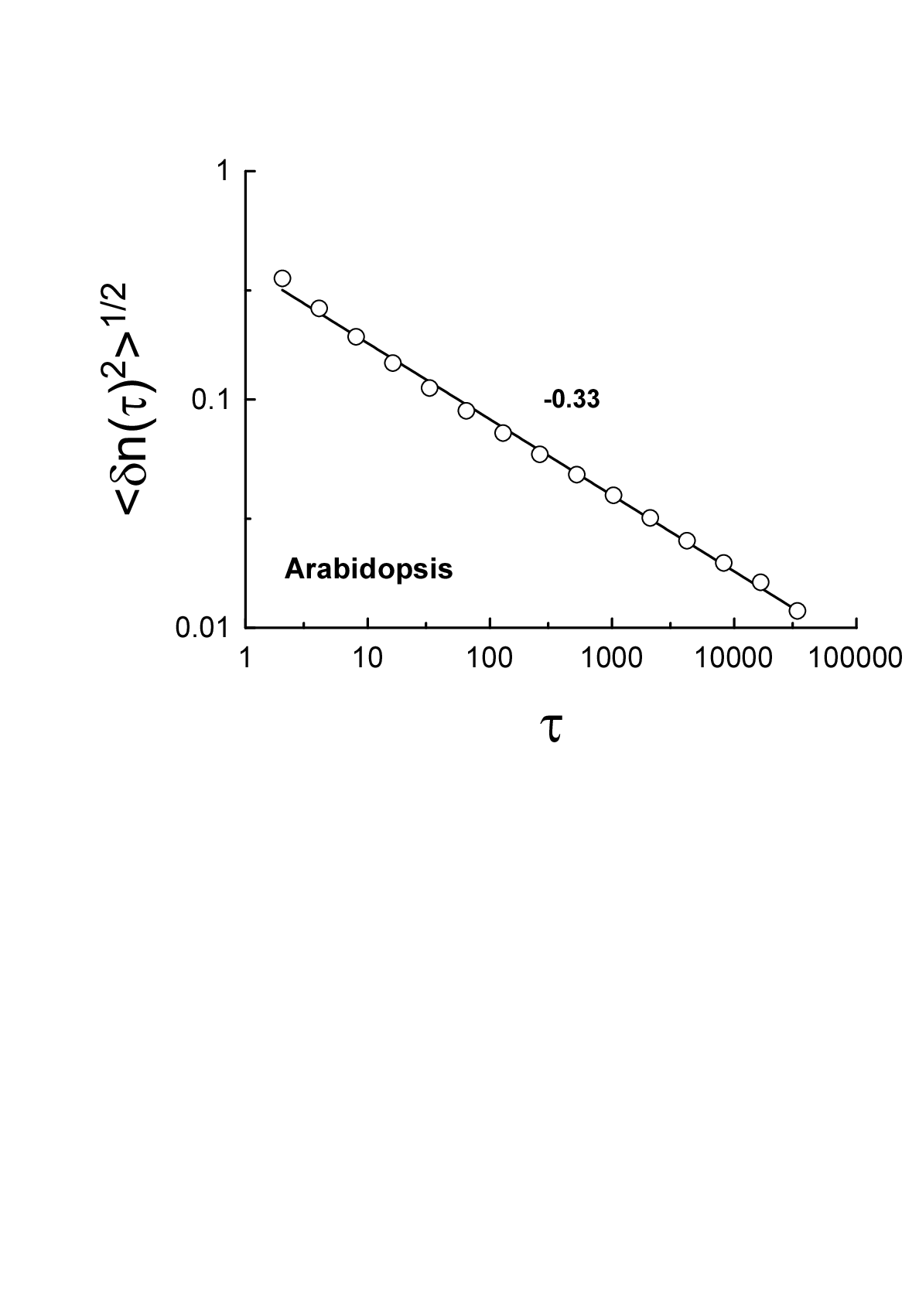} \vspace{-5cm}
\caption{The standard deviation for $\delta n (\tau)$ vs $\tau$ for
T-dominated sub-sequence of the Arabidopsis (in log-log scales). The straight
line (the best fit) indicates the scaling law Eq. (2).}
\end{figure}

 Because of many orders of space scales involved in these processes 
one can expect that the clustering will exhibit scale-invariant properties (see, for instance, 
Ref. \cite{stan1}). A cluster-scaling for 
stochastic systems was recently suggested in Refs. \cite{sb},\cite{b2},\cite{b3}. 
The genome data can be readily checked on the cluster-scaling properties in a "1 or 0" mapping. 
In this presentation (see, for instance Refs. \cite{voss},\cite{pod} and references 
therein) one should put A=1 and C=G=T=0 in an original DNA sequences to obtain an A-dominated sub-sequences 
(one can obtain C or G, or T-dominated sub-sequences in analogous way). Then, to study statistical 
clustering in sub-sequences $\left\{a_i\right\}$ (where $a_i$=1 or 0 and $i=1,2...$) one should take running average:
$$
n_j(\tau) = \frac{1}{\tau} \sum_{i=j}^{i=j+\tau} a_i   \eqno{(1)}
$$   
along the sub-sequences.
For the 1 or 0 mapping this running average will present a weight of the sub-sequences in 
interval [$j$,$j+\tau$]. Following to Ref. \cite{sb} we are interested in scaling 
variation of the standard deviation of the running density fluctuations $\langle \delta
n_j(\tau)^2 \rangle^{1/2}$ with $\tau$
$$
\langle \delta n_j(\tau)^2 \rangle^{1/2} \sim \tau^{-\alpha}
\eqno{(2)}
$$
where $\langle...\rangle$ denotes average over the sub-sequences, $\delta n_j(\tau)= n_j(\tau) - 
\langle n (\tau) \rangle$. The power law, Eq. (2), corresponds to a scale-invariant (scaling) behavior. 
\begin{figure} \vspace{-1cm}\centering
\epsfig{width=.45\textwidth,file=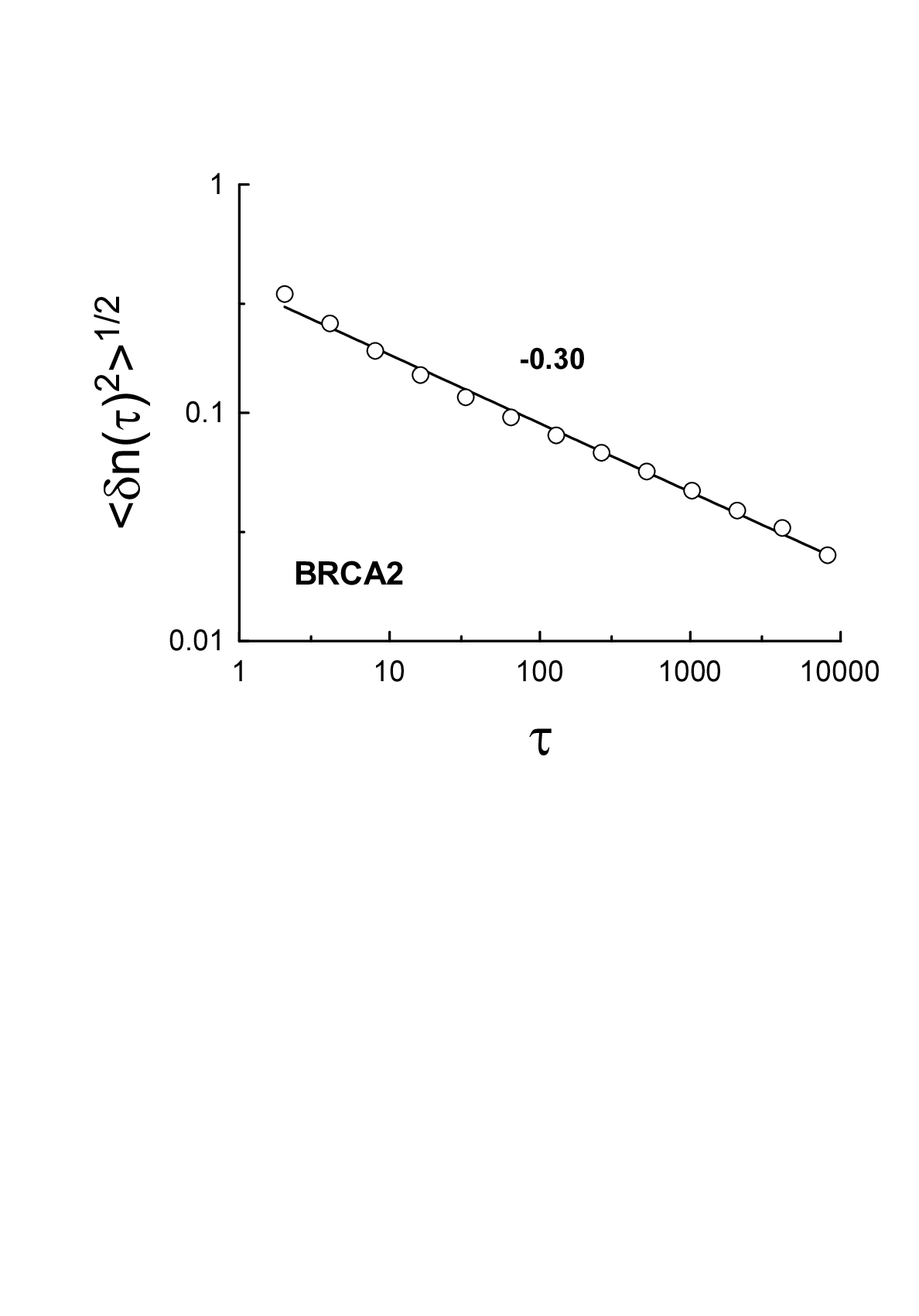} \vspace{-5cm}
\caption{The standard deviation for $\delta n (\tau)$ vs $\tau$ for
T-dominated sub-sequence of gene BRCA2 (in log-log scales). The straight
line (the best fit) indicates the scaling law Eq. (2).}
\end{figure}

The exponent $\alpha$ in Eq. (2) was called in Ref. \cite{sb} as cluster-exponent. 
For white noise zeros (intersections of a white noise signal with time axis) it can be derived analytically that $\alpha = 1/2$ (see Ref. \cite{sb} and references therein). This value can be considered as an upper limit 
(non-clustering case) for the cluster-exponent. If $0< \alpha < 0.5$ we have a cluster-scaling situation, and 
the cluster-scaling is stronger for smaller values of $\alpha$ (see for examples Ref. \cite{sb}).

In this paper we will present, as an example, results obtained for the genome sequence of the flowering plant Arabidopsis thaliana, which is a model plant for genome analysis \cite{mei} and for two human genes BRCA2 and NRXN1. 

Let us start from the Arabidopsis. Its genome is one of the smallest plant genomes  
(about 157 million base pairs and five chromosomes) that makes Arabidopsis thaliana useful for genetic 
mapping and sequencing. The most up-to-date version of the Arabidopsis thaliana genome is maintained 
by The Arabidopsis Information Resource (TAIR) (see, for instance, 
http://www.plantgdb.org/). 
The results of computations for the genome sequences associated with the Arabidopsis are shown in figure 2. 
We show in Fig. 2 results for the T-dominated sub-sequence, whereas the results for A, C, and G-dominated sub-sequences are similar to those shown in the Fig. 2. The Fig. 2 shows (in the log-log scales) dependence of the 
standard deviation of the running density fluctuations $\langle \delta n (\tau)^2 \rangle^{1/2}$ on
$\tau$ for the T-dominated subsequence. The straight line is drawn in this figure to indicate the
scaling (2). The slope of this straight line provides us with the cluster-exponent  $\alpha = 0.33 \pm 0.02$. 
\begin{figure} \vspace{-1cm}\centering
\epsfig{width=.45\textwidth,file=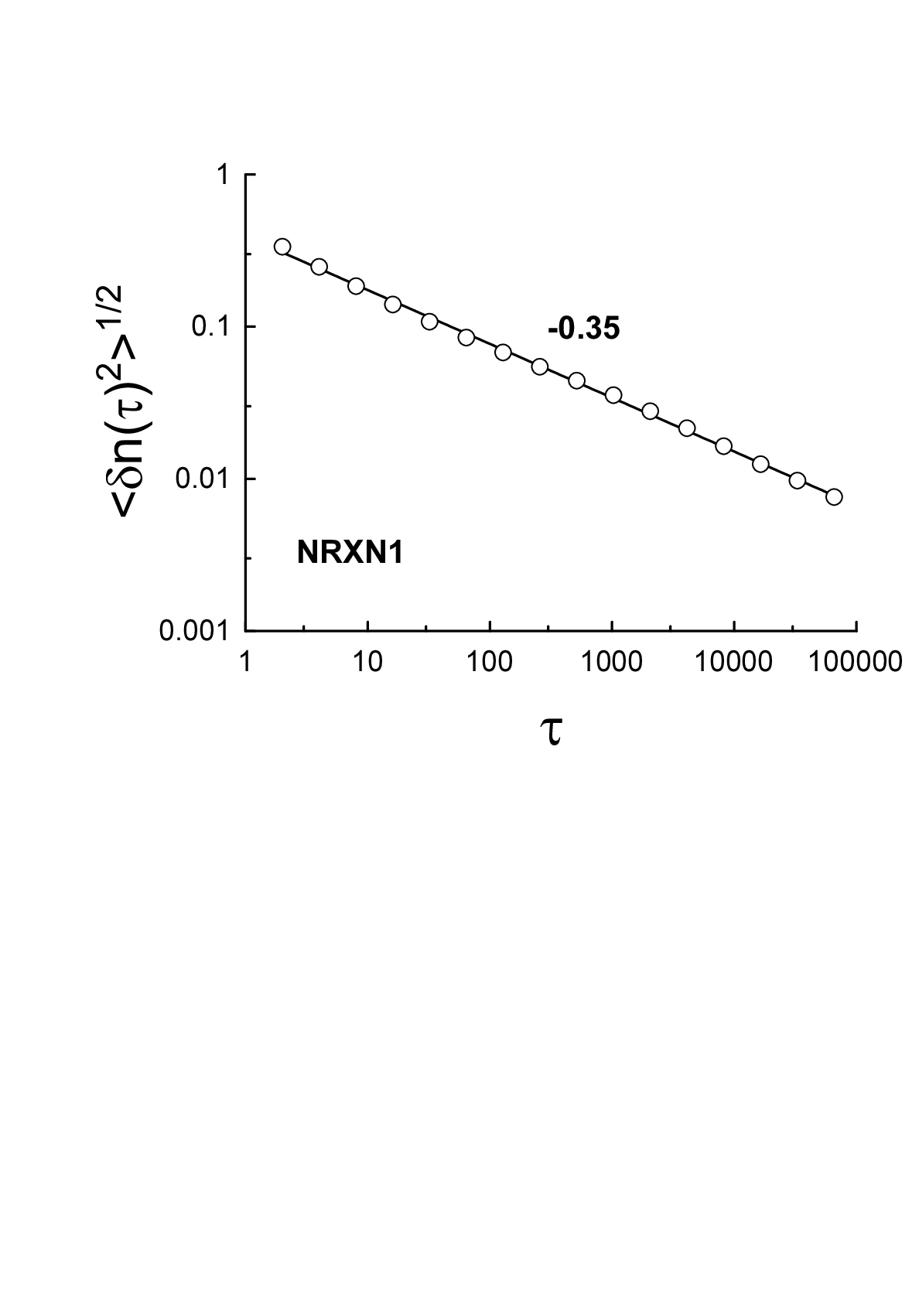} \vspace{-5cm}
\caption{The standard deviation for $\delta n (\tau)$ vs $\tau$ for
T-dominated sub-sequence of gene NRXN1 (in log-log scales). The straight
line (the best fit) indicates the scaling law Eq. (2). }
\end{figure}
  The results of computations for the genome sequences associated with genes: BRCA2 and NRXN1, 
are shown in figures 3 and 4 respectively (the full set of the genome sequences can be found in site: 
http://www.ncbi.nlm.nih.gov). Molecular location of gene BRCA2 on chromosome 13: base pairs 
32,889,616 to 32,973,808). BRCA2 gene helps prevent cells from growing and dividing too rapidly or 
in an uncontrolled way. By helping repair DNA, BRCA2 plays a role in maintaining the stability 
of a cell's genetic information. Gene NRXN1 (neurexin 1) is among the largest known in human,
molecular location on chromosome 2: base pairs 50,145,642 to 51,259,673.
NRXN1 gene represents a strong candidate for involvement in the etiology of nicotine dependence, and even subtle changes in NRXN1 might contribute to susceptibility to autism. \\
\begin{figure} \vspace{-1.2cm}\centering
\epsfig{width=.45\textwidth,file=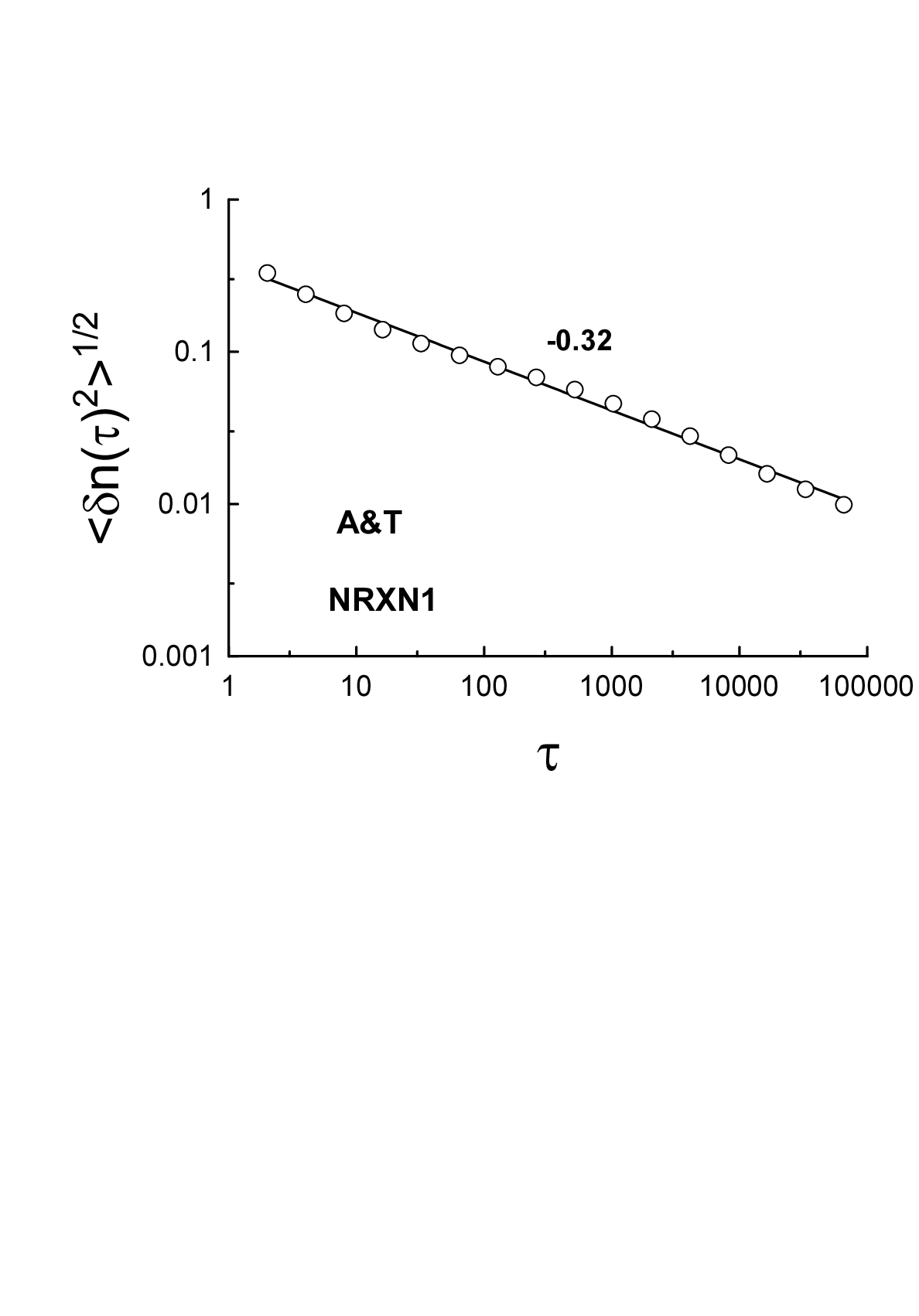} \vspace{-4.8cm}
\caption{The standard deviation for $\delta n (\tau)$ vs $\tau$ for
A\& T (circles) dominated sub-sequence of gene NRXN1
(in log-log scales). The straight lines (the best fit) indicate the scaling law Eq. (2). }
\end{figure}

We show in Figs 3 and 4 results for the T-dominated sub-sequences, 
whereas the results for A, C, and G-dominated sub-sequences are similar to those shown in the 
Figs. 3 and 4 (for each gene respectively). Fig. 3 shows (in the log-log scales) dependence of the 
standard deviation of the running density fluctuations $\langle \delta n (\tau)^2 \rangle^{1/2}$ on
$\tau$ for the T-dominated subsequence of gene BRCA2. The straight line is drawn in this figure to indicate the
scaling (2). The slope of this straight line provides us with the cluster-exponent  $\alpha = 0.30 \pm 0.02$. 
Figure 4 shows analogous result for gene NRXN1 with $\alpha = 0.35 \pm 0.02$. One can see that in both cases 
we have rather strong cluster-scaling with the cluster-exponent different for the different genes. 
The main consequence of the finite-size effects for the cluster-scaling is a wavy character of the scaling 
data in the log-log scales (cf. Ref. \cite{sb}).  

\section{Hydrogen bonds} 

The most popular potential for modeling the hydrogen (H) bond within a base-pair in the DNA chains is the 
Morse potential (see, for instance, Refs. \cite{pb},\cite{dpb},\cite{cam},\cite{hen}):
$$
V_i (y_i) = D_i\left[ \exp-\left(\frac{a}{2} y_i \right) -1\right]^2  \eqno{(3)} 
$$
where $D_i$ is the site-dependent dissociation energy of the $i$th pair, which can take two values $D_i=D^{A-T}$ and 
$D_i= D^{C-G}$ for the A-T and the C-G pairs in the $i$th site respectively (the A-T pair includes two H bonds, 
while the C-G pair includes three H bonds, see Introduction); $a^{-1}$ is a measure of the potential well
width; variable $y_i$ is a dynamical deviation of the H bonds from their equilibrium lengths at position $i$. 
The ratio $D^{C-G}/D^{A-T}=1.5$ is often used for the model 
purposes (though recent quantum chemical calculations \cite{spon} results in a ratio 
$D^{C-G}/D^{A-T}=2$).

Randomly distributed along the DNA chain bivalued H-bond coupling
strengths $D^{A-T}$ and $D^{G-C}$ are usually used in the DNA dynamics models. This would be appropriate 
for an arbitrary base pair sequence. However, as it is follows from previous 
consideration, homogeneous random distribution is not realistic even for the most long genes like 
NRXN1 (see, for instance, Ref. \cite{li}). 
The dynamic heterogeneous properties of DNA molecule was considered, for instance, as a 
reason for the so-called multi-step melting \cite{ch}. However, the assumption of a random 
and short-range (delta-) correlated sequence made in Ref. \cite{ch} do not result in the multi-step 
melting and only an additional assumption of an additional backbone stiffness due to the double-stranded 
conformation of DNA molecule allowed to the authors to observe a multi-step melting in their model. 
In the model suggested In Ref.  \cite{je} the sequence randomness considered as a quenched noise with 
finite sequence correlation length. In this approach regions dominated by A-T or, alternatively, by C-G pairs 
play significant role in the bubble (i.e. locally denaturated states) formation. \\
\begin{figure} \vspace{-1.8cm}\centering
\epsfig{width=.45\textwidth,file=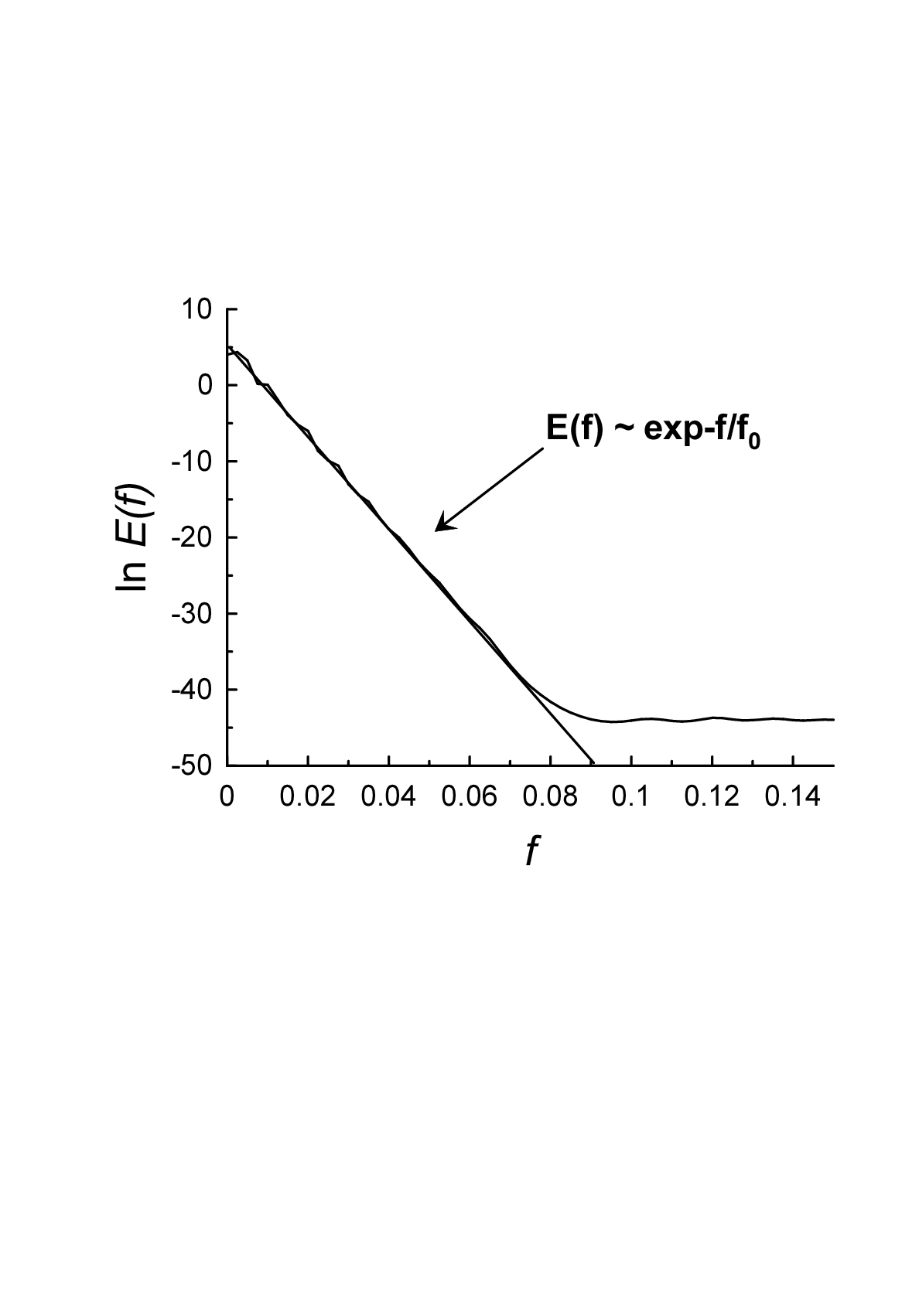} \vspace{-3.8cm}
\caption{Spectrum of a chaotic solution of the Duffing oscillator. The solid straight line indicates exponential spectrum. }
\end{figure}
\begin{figure} \vspace{-0.5cm}\centering
\epsfig{width=.45\textwidth,file=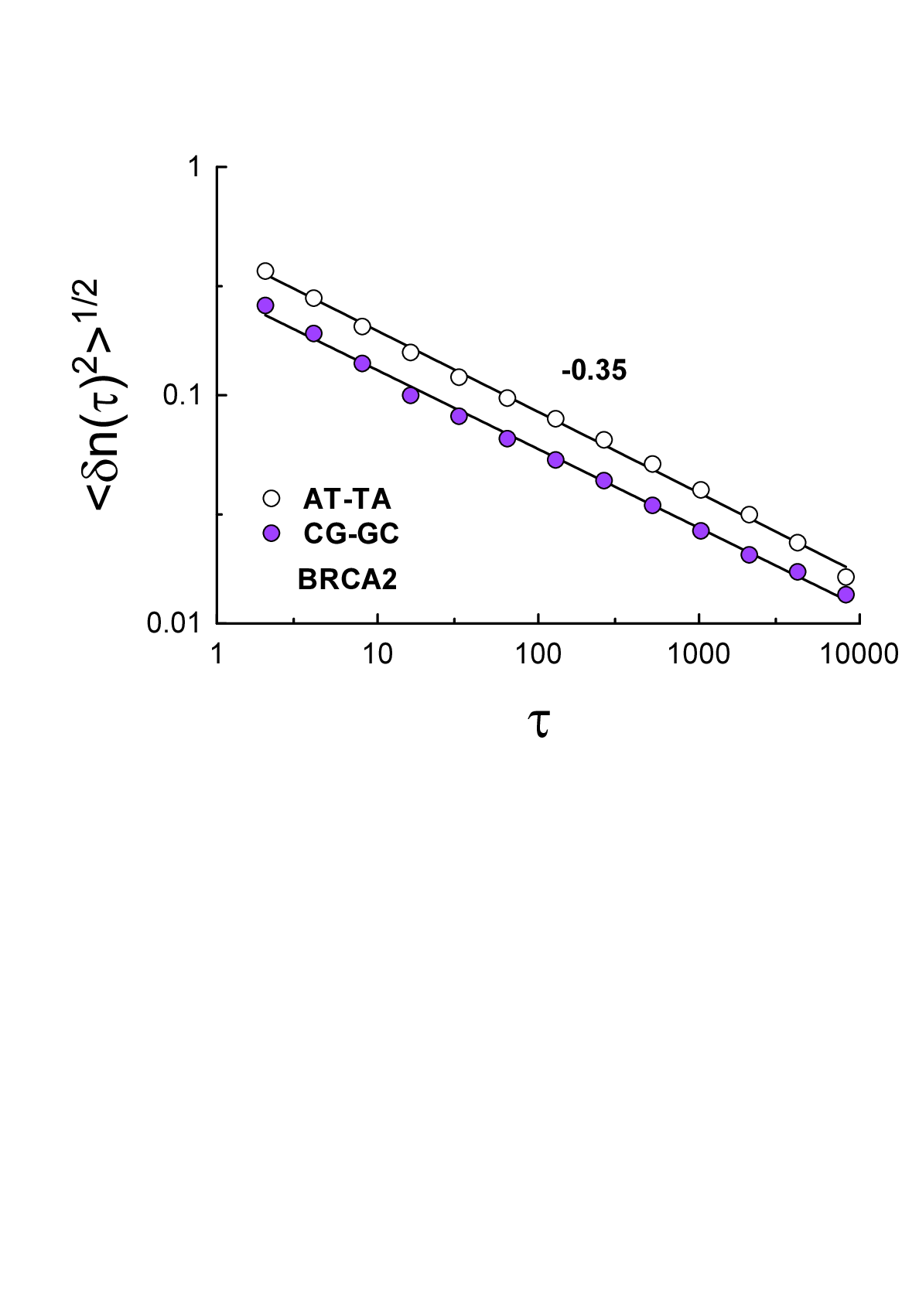} \vspace{-5cm}
\caption{The standard deviation for $\delta n (\tau)$ vs $\tau$ for
AT-TA (circles) and CG-GC (filled circles) dominated sub-sequence of gene BRCA2 
(in log-log scales). The straight lines (the best fit) indicate the scaling law Eq. (2). }
\end{figure}
Taking into account the cluster-scaling of the DNA nucleotides is a natural step toward more realistic dynamical 
model. Because of the bivalued H-bond coupling strengths: $D_i=D^{A-T}$ or $D_i=D^{G-C}$, this can be readily 
done using following bivalued mapping: $A=T=1,~ C=G=0$ or, alternatively, $C=G=1,~ A=T=0$. 
Figure 5 shows cluster-scaling behavior, Eq. (2), for the former mapping of NRXN1 gene. The cluster scaling exponent $\alpha = 0.32 \pm 0.02$ in this case. For $C=G=1,~ A=T=0$ the mapping calculations give the same result 
as well as for corresponding mappings of the gene BRCA2 (indication of an universality). Therefore, the bivalued sequences of the $D_i$ coefficients for the DNA dynamic chain should be chosen as cluster-scaling ones with certain 
cluster-exponent $\alpha$ Eq. (2) (for the considered genes $\alpha \simeq 0.32$).    \\

  As it is mentioned in a recent Ref. \cite{em} the systems with the Morse potential are too complex for analysis because the exponential form of the potential. Therefore it is suggested in the Ref. \cite{em} to use the Duffing potential instead:
$$
V_i (y_i) = \beta y_i^2+\alpha y_i^4  \eqno{(4)}
$$
where $\alpha$ and $\beta$ are constants. The authors of the Ref. \cite{em} claim that these potentials have similar behaviour but the systems with the Duffing potential much easier to analyse. In the Ref. \cite{ss} the Duffing potential is suggested for stacking interactions (see next Sections) on the basis of single molecule experiments. 

  The Duffing oscillator 
$$
\ddot{x} + b\dot{x} +\beta x + \alpha x^3 = A \sin  \omega t  \eqno{(5)}
$$  
is a driven oscillator with a non-linear elasticity (at $b=0$ it is a Hamiltonian system). At certain values of the parameters this oscillator exhibits chaotic behaviour (see, for instance, Ref. \cite{spot}). 
Figure 6 shows power spectrum for this oscillator for a chaotic set of the parameters \cite{spot}: $b= 0.05$, $\beta = -1$, $\alpha =1$, $A= 0.7$ and $\omega = 0.7$. The solid straight line indicates exponential type of spectrum (this is relevant for the next Sections, cf. Fig. 10).  

\section{Stacking interaction}

\begin{figure} \vspace{-1.5cm}\centering
\epsfig{width=.45\textwidth,file=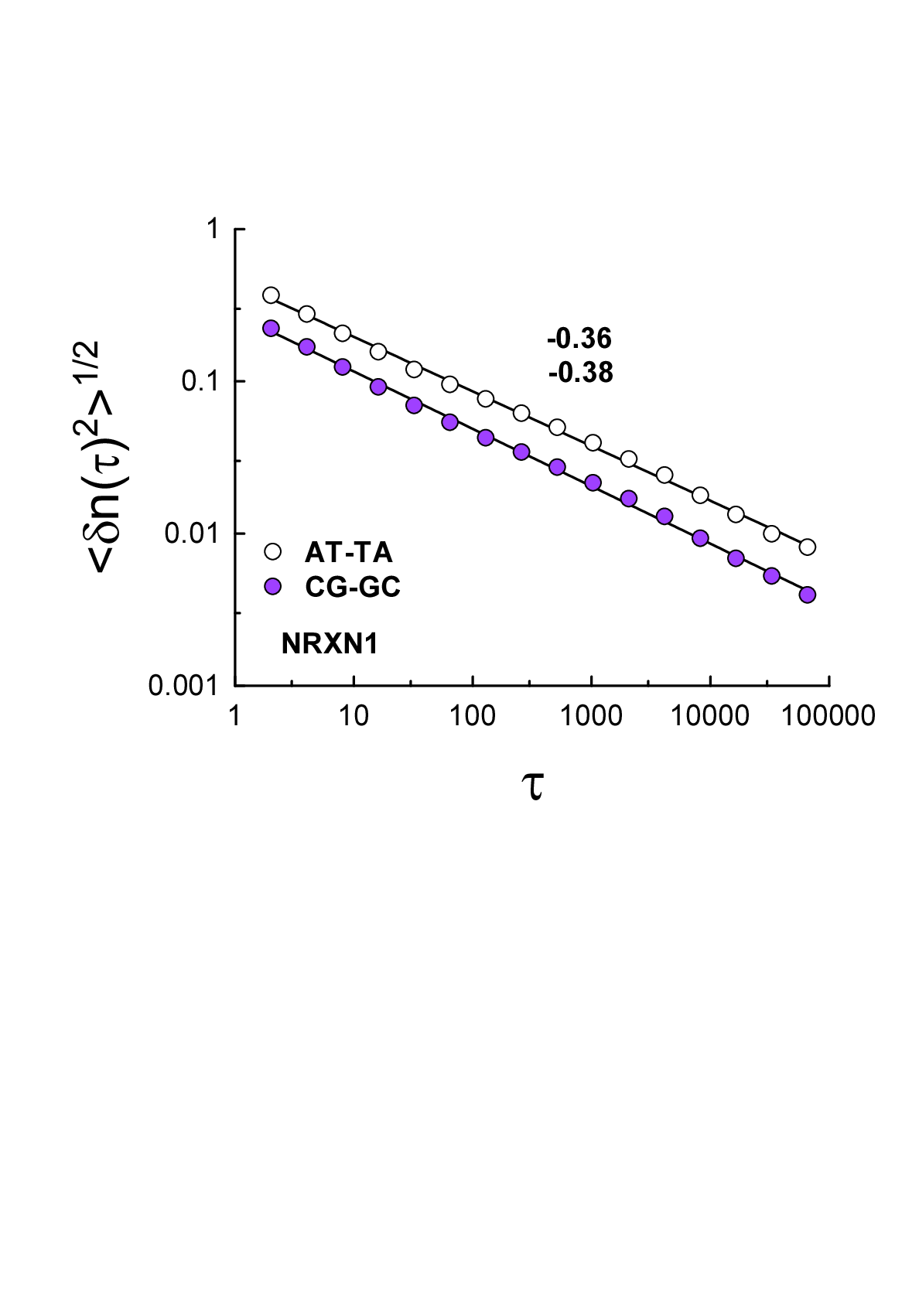} \vspace{-4.5cm}
\caption{The standard deviation for $\delta n (\tau)$ vs $\tau$ for
AT-TA (circles) and CG-GC (filled circles) dominated sub-sequence of gene NRXN1
(in log-log scales). The straight lines (the best fit) indicate the scaling law Eq. (2). }
\end{figure}
Two factors are mainly responsible for the stability of the DNA double helix: 
base pairing between complementary strands and stacking between adjacent
bases (see Introduction). It is shown experimentally that DNA stability is mainly determined by
base-stacking interactions which contribute greatly into the dependence of the duplex 
stability on its sequence. (see, for instance, Ref. \cite{yak}). 
Therefore, it is interesting to check whether the stacking interactions dominate also the above-considered cluster-scaling phenomenon (cf. Introduction). In order to check this let us use following mapping: in combination $AT=TA=1~1$, 
and $A=T=G=C=0$ otherwise. An alternative mapping is: in combination $CG=GC=1~1$, and $A=T=G=C=0$ 
otherwise. If the stacking interactions dominate the cluster-scaling phenomenon, then one can expect 
that the cluster-scaling will be more pronounced just for these maps (cf. Fig. 1). It means that 
the cluster-exponent corresponding to these maps would be {\it smaller} than cluster-exponents 
observed for the above considered maps. As one can see 
comparing  Figs. 7 and 8 with Figs. 3,4, and 5 in reality we have an opposite situation. 
This comparison indicates that the stacking interactions 
do not dominate the above-considered cluster-scaling phenomenon (at least for the examples given 
in the paper).

In a realistic dynamic model of DNA molecule one should take into account also the cluster-scaling 
of stacking interaction itself as it is shown in Figs. 7 and 8 for instance (see also Refs. \cite{am},\cite{kru} 
for heterogeneity of both pairing and stacking interactions). 
This can be done in the frames of a commonly used approximation for the stacking potential (see, for instance, 
Ref. \cite{jf}) 
$$
W_i (y_i,y_{i-1}) = \frac{\Delta H_i}{C} \left(1-\exp\left(-b(y_i-y_{i-1})^2\right)\right)  \eqno{(6)}
$$
where $\Delta H_i$ can take different values for different stacked pairs $\left\{y_i,y_{i-1}\right\}$. 
Because the situation is not bivalued in this case this task seems to be more difficult than for the hydrogen bonds. 
The main problem here is hybridization of the nucleotides in different types of the stacked base-pairs (Fig. 1). 
The fact that the cluster-scaling exponents for the different types of stacked base-pairs have approximately the same 
value can help to solve this problem. This is not the case, however, for hybridization problem if one will 
consider a realistic model taking into account cluster-scaling of both hydrogen bonds and stacking interactions 
(the cluster-scaling exponents are different for hydrogen bonds: Fig. 5, and for stacking interactions: Figs. 7 and 8).

\section{Chaotic order of the stacking interactions}

Both stochastic and deterministic processes can result in the
broad-band part of the spectrum, but the decay in the
spectral power is different for the two cases. An exponential decay with respect 
to frequency refers to chaotic time series while a power-law decay indicates that the spectrum 
is stochastic. Not all chaotic systems have the exponentially decaying spectra, but 
appearance of an exponentially decaying spectrum in the system under consideration 
provides a strong indication that we have deal with a chaotic (deterministic) process \cite{sig}-\cite{fm}.
\begin{figure} \vspace{-1cm}\centering
\epsfig{width=.45\textwidth,file=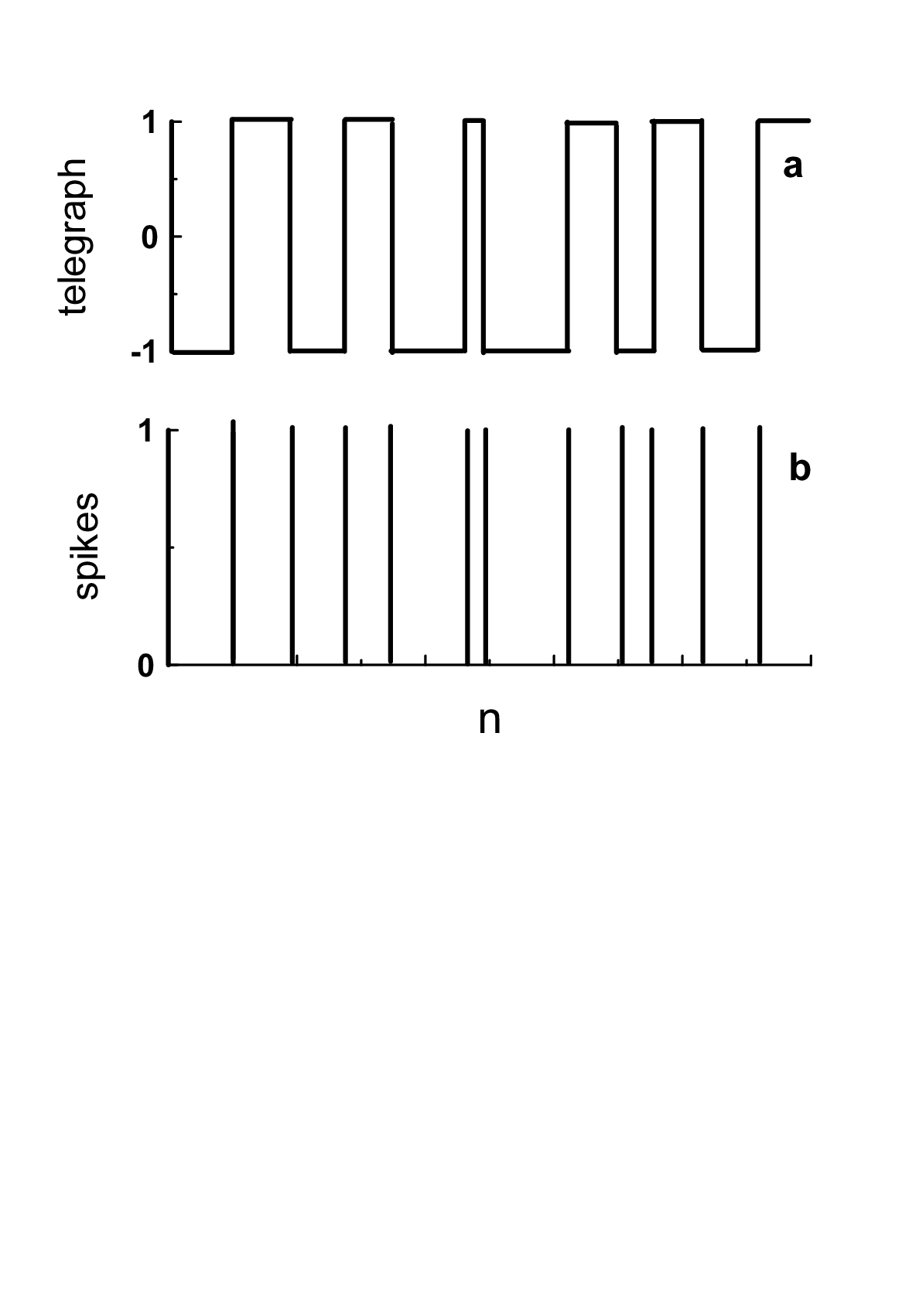} \vspace{-5cm}
\caption{Mapping of a spiky sequence (figure 9b) into a telegraph one (figure 9a).  }
\end{figure}
The previously observed spectra of the DNA-sequences mappings exhibited power-law decay indicating 
a stochastic origin of the DNA-sequences randomness \cite{voss}-\cite{arn}. It should be noted that 
all the maps used in these investigations operated with simple use of the $A$ or/and $T$ or/and $G$ or/and $C$ 
numerical mapping. It seems, however, that a deeper insight in the underlying physics can be 
obtained using numerical maps operating with the combinations AT and TA which represent energy minima for the 
stacking interactions (cf. Fig. 1 and Ref. \cite{b2}). For this purpose we will put combinations 
$AT=TA=1$, and $A=T=G=C=0$ otherwise in a DNA-sequence under consideration (the multiple AT/TA 
combinations will be also considered as a single '1' in this mapping, for example: ATATTA=1). This map will 
represent a \{0,1\}-values map of the stacking interactions energy minima sequence. 
\begin{figure} \vspace{-0.7cm}\centering
\epsfig{width=.45\textwidth,file=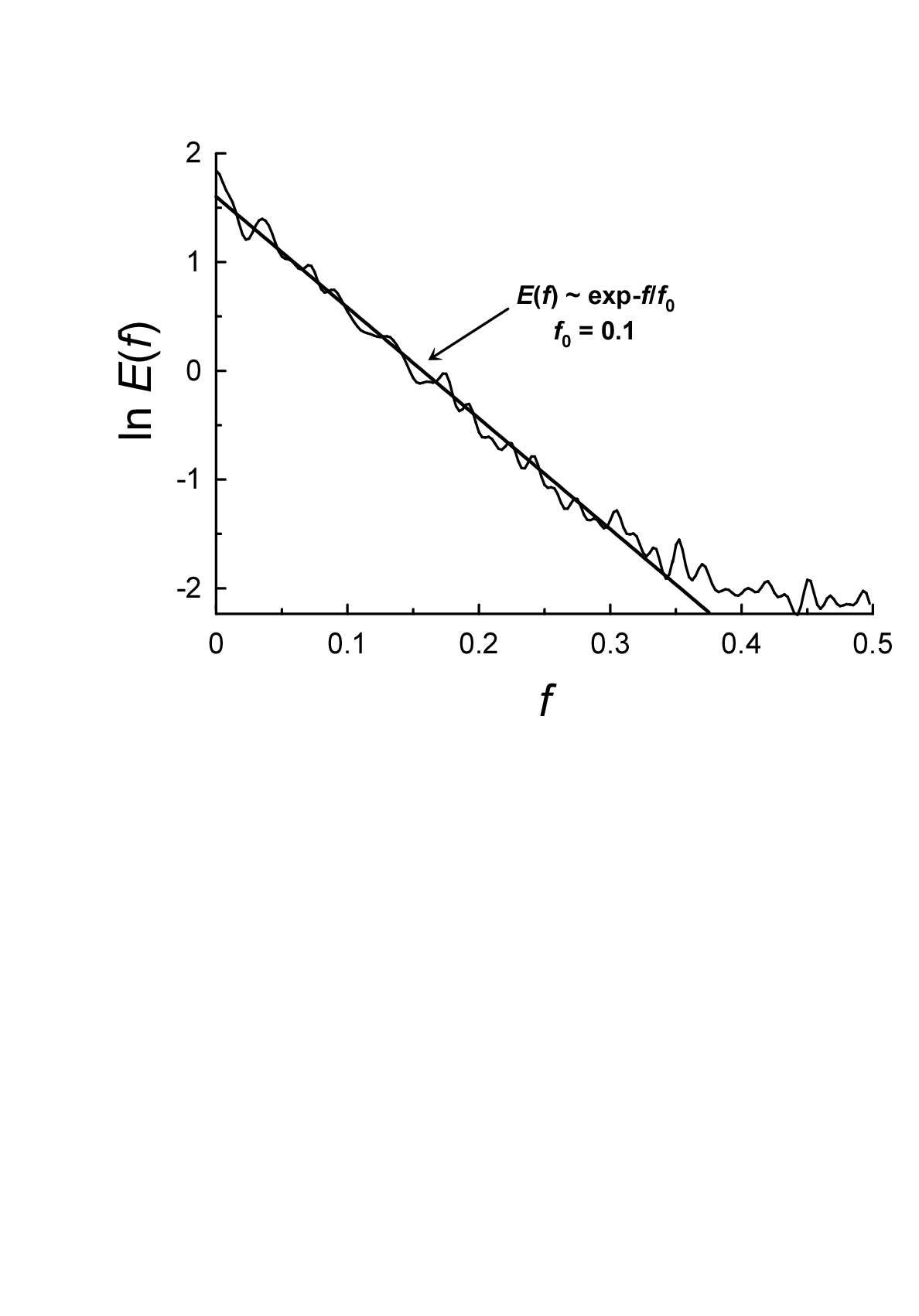} \vspace{-5cm}
\caption{Spectrum of a telegraph series corresponding to the energy minima map 
constructed for the genome sequence of the Arabidopsis thaliana.  }
\end{figure}

An additional technical problem will appear when one will try to analyze spectral properties of 
such map. The sequence will be very spiky and the usual spectral methods (such as the fast Fourier 
transform or the maximum entropy method, for instance) will be practically useless. In order to solve 
this problem we will use an additional mapping of the spiky series into a telegraph signal. 
The spikes (symbols 1) are identical to each other and the dynamical information 
is coded in the length of the interspike intervals and the interspike intervals positions 
on the sequence, therefore it is the most direct way to map the spiky sequence 
into a telegraph signal, which has values -1 from one side of a spike and values +1 from 
another side of the spike. An example of such mapping is given in figure 9. 
While the dynamical information is here the same as for the corresponding spike sequence, 
the spectral methods are quite applicable to analysis of the telegraph series.   

  Figure 10 shows spectrum of a telegraph sequence corresponding to the above-described map 
constructed for the genome sequence of the Arabidopsis thaliana. We used the semi-log 
axes in the Fig. 10 in order to indicate exponential decay of the spectrum (the straight line, $f_0 \simeq 0.1$).

Many of the well known chaotic attractors ('Lorenz', 'R\"{o}ssler', etc.) exhibit the exponentially decaying spectra \cite{o}. In the Ref. \cite{ss} the Duffing potential is suggested just for stacking interactions, instead of the Eq. (6) (see Fig. 6). Here, for comparison with the Fig. 10, we will consider also a chaotic spectrum generated by 
the Kaplan-Yorke map \cite{ky} (relevance of this choice will be clear immediately). In the Langevin 
approach to Brownian motion the equation of motion is 
$$
\dot{y}= -\gamma y + F(t)  \eqno{(7)}
$$
where the fluctuating kick force on the particle is a Gaussian white noise: 
$F(t)= \sum_n \eta_n \delta (t-n\tau)$ and $y(t)$ to take values in $R^m$.
One can assume \cite{beck} that the evolution of the kick strengths 
is determined by a discrete time dynamical system $T$ on the phase space and projected onto $R^m$ 
by a function $f$:
$$
\eta_n = f(x_{n-1}),~~~~  x_{n+1}= T x_n  \eqno{(8)}
$$

\begin{figure} \vspace{-1.6cm}\centering
\epsfig{width=.45\textwidth,file=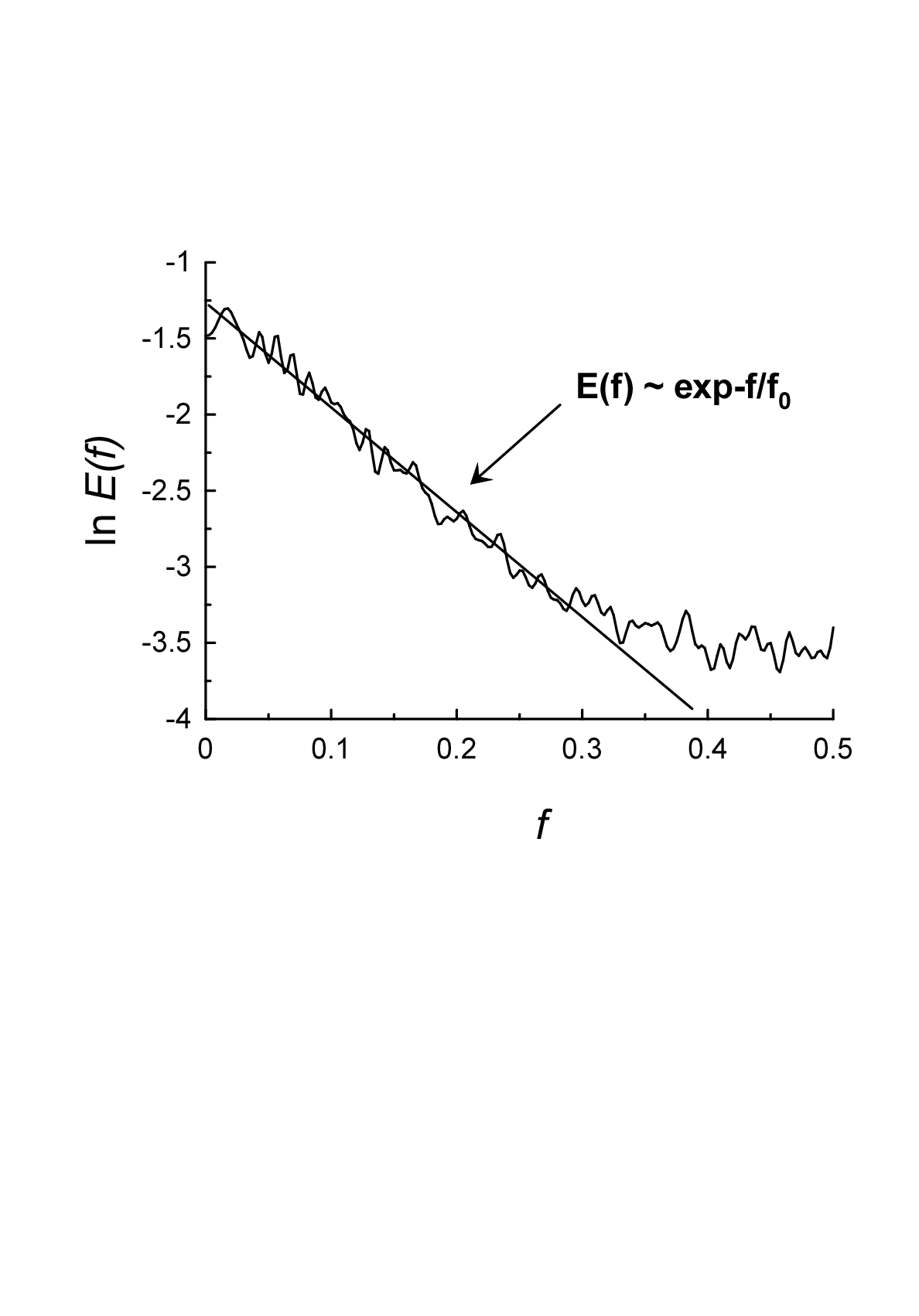} \vspace{-4cm}
\caption{Spectrum of a chaotic solution of the Kaplan-Yorke map. }
\end{figure}
Then the solution of Eq. (7) is
$$
y(t) = e^{-\gamma(t-n\tau)} y_n \eqno{(9)}
$$
where $n$ equals the integer part of the relation $t/\tau$ and the recurrence
$$
x_{n+1}= T x_n, ~~~ y_{n+1}=\alpha y_n + f(x_n)  \eqno{(10)}
$$
provides value of $y_n$ (with $\alpha = e^{-\gamma\tau}$). In certain sense the dynamical
system (8) is equivalent to the stochastic differential equation (7). In the 
generalization related to the Eq. (10) the force $F(t)$ can be considered as a {\it non}-Gaussian 
process which is determined by $f$ and $T$. The Kaplan-Yorke map \cite{ky},\cite{jo},\cite{spot} is a particular 
simple case for this generalization:
$$
Tx = 2x~ (mod~ 1), ~~~ f(x)= \cos 4 \pi x  \eqno{(11)}
$$
Figure 11 shows spectrum of a chaotic solution of the Kaplan-Yorke map ($\alpha =0.2$). We used the semi-log 
axes in this figure in order to indicate exponential decay of the spectrum (the straight line). 

\begin{figure} \vspace{-0.8cm}\centering
\epsfig{width=.45\textwidth,file=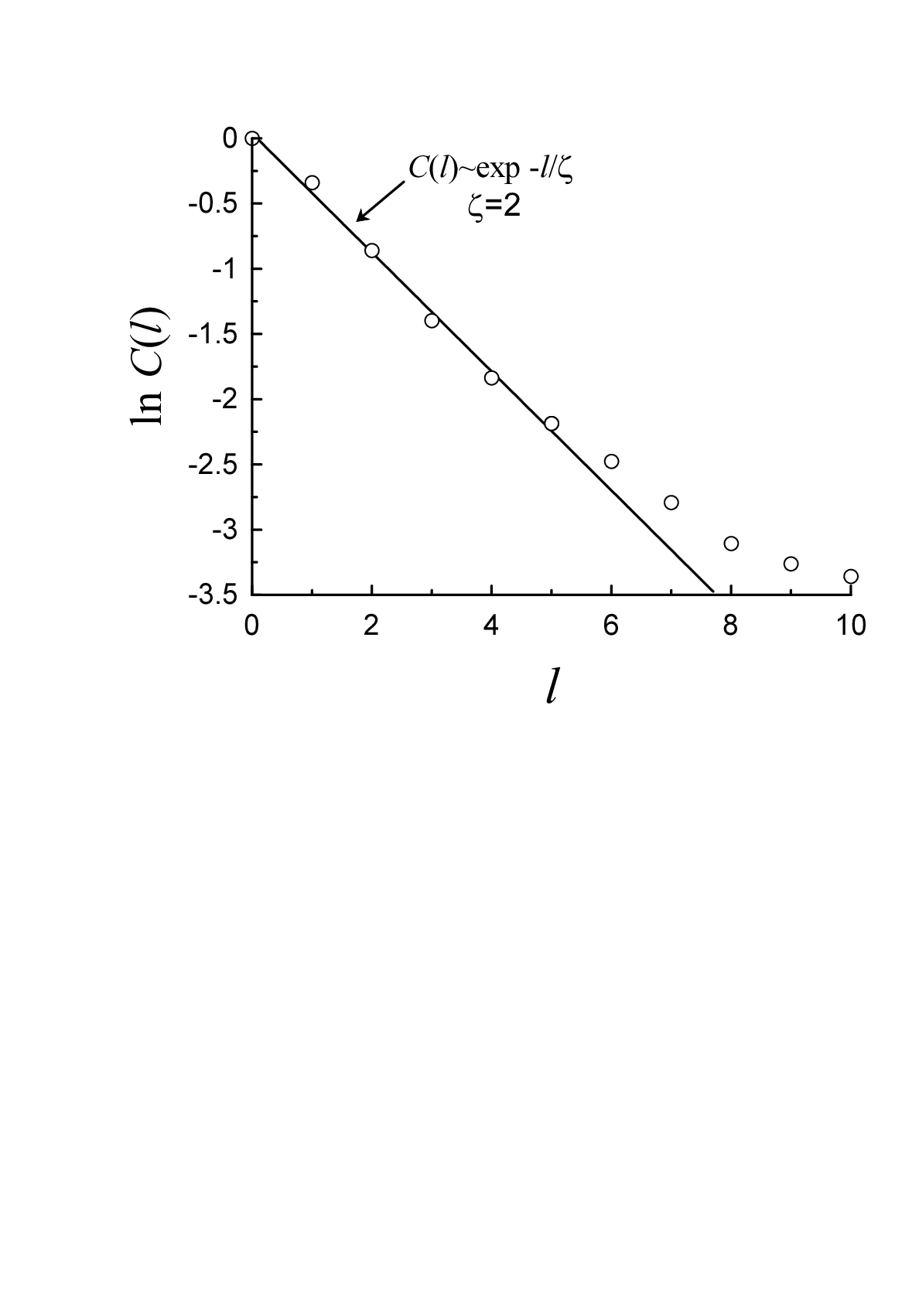} \vspace{-5cm}
\caption{Autocorrelation function corresponding to the spectrum shown in Fig. 10.  }
\end{figure}
Although the exponential part of the spectrum in Fig. 10 is apparently extended to the frequencies $f \simeq 0.3$ 
for frequencies larger then $f\simeq 0.2$ (i.e. for scales $n \leq 5$) corresponding telegraph signal 
is a random one. This can be seen from the figure 12, which shows autocorrelation function corresponding 
to the spectrum shown in Fig. 10 (the correlation length $\zeta =2$). We have a chaotic order on the 
large scales only (see also next section, Fig. 14). Fig. 13 shows analogous autocorrelation function for 
the Kaplan-Yorke solution.

\section{Large-scale chaotic coherence}
\begin{figure} \vspace{-0.5cm}\centering
\epsfig{width=.45\textwidth,file=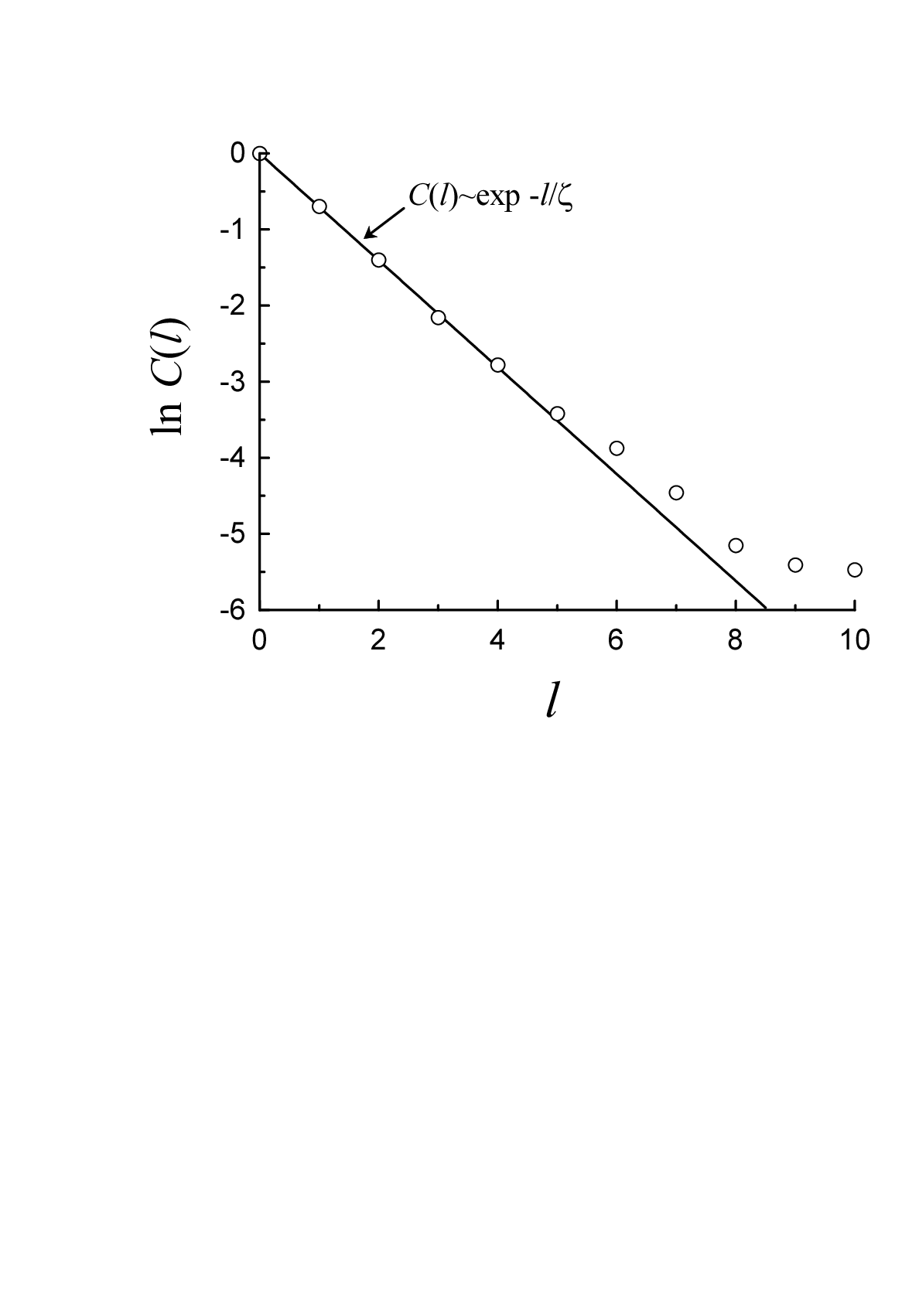} \vspace{-5cm}
\caption{Autocorrelation function corresponding to the spectrum shown in Fig. 11.  }
\end{figure}
In the double-stranded DNA the two strands are complementary in a local sense, i.e. the nucleotide bases 
pair up such that A always pairs with T and G always pairs with C. But what can one say about nonlocal 
coherence of the nucleotides' sequences in the two strands? Actually, because of the local complimentary 
behavior this question can be answered by studying coherence between the A (map: A=1, T=C=G=0) and T 
(map: T=1, A=C=G=0) dominated sequences along a single strand (analogously for the C and G sequences). 
Due to the complementary properties of the A and T nucleotides the chaotic (deterministic) order of the 
energy minima of the stacking interactions should result in a large-scale coherence of the two DNA strands' 
sequences. This is mainly relevant to A and T sequences because just the AT/TA compositions correspond to 
the energy minima. Certain (however considerably smaller) large-scale coherence can appear also in C 
and G sequences as a secondary effect 
(see Fig. 14 and the explanations below).   
\begin{figure} \vspace{-0.9cm}\centering
\epsfig{width=.45\textwidth,file=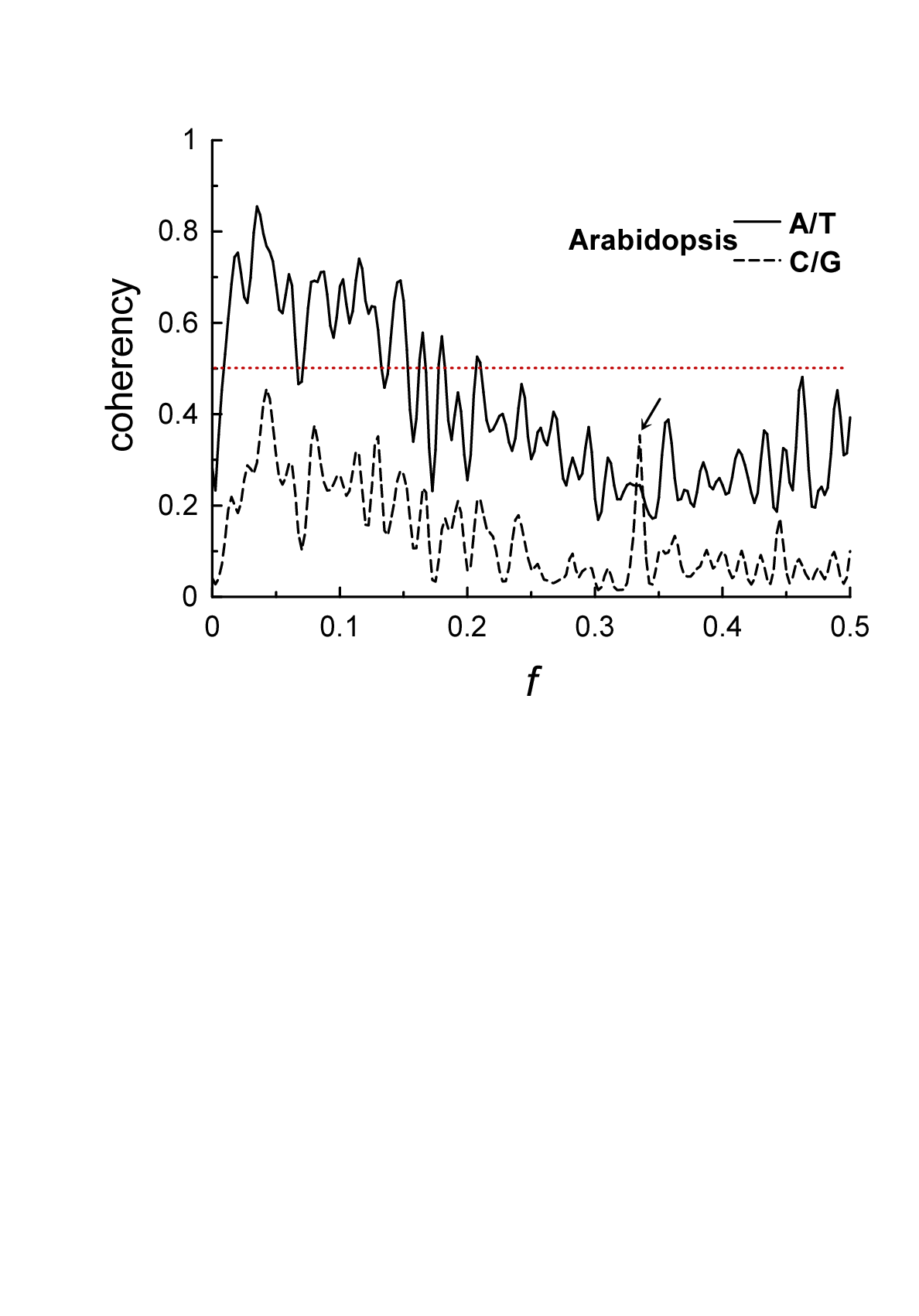} \vspace{-4.87cm}
\caption{Coherency of the two DNA strands' sequences  for the Arabidopsis thaliana.}
\end{figure}
In order to compare coherent properties of the two DNA strands' sequences we will use 
cross-spectral analysis. The cross spectrum $E_{1,2}(f)$ of two processes $x_1(t)$ and $x_2(t)$ is defined 
by the Fourier transformation of the cross-correlation function normalized by the product of square root of the univariate power spectra $E_1(f)$ and $E_2(f)$:
$$
E_{1,2}(f)= \frac{\sum_{\tau} \langle x_1(t)x_2(t-\tau) \rangle 
\exp (-i2\pi f \tau)}{2\pi \sqrt{E_1(f)E_2(f)}} \eqno{(12)}
$$
the bracket $\langle... \rangle$ denotes the expectation value. The cross spectrum can be decomposed
into the phase spectrum $\phi_{1,2} (f)$ and the coherency $C_{1,2}(f)$:
$$
E_{1,2}(f)= C_{1,2}(f) e^{-i \phi_{1,2} (f)}  \eqno{(13)}
$$
Because of the normalization of the cross spectrum the coherency
is ranging from $C_{1,2}(f)=0$, i.e. no linear relationship between
$x_1(t)$ and $x_2(t)$ at $f$, to $C_{1,2}(f)=1$, i.e. perfect linear relationship. 
 
Figure 14 shows coherency of the A (or T) dominated sequences on the two strands of 
the Arabidopsis DNA-duplex (solid curve), and coherency of the C (or G) dominated sequences (dashed curve). 
While the A (or T) dominated sequences exhibit rather high ($> 0.5$) coherency in  a low-frequency domain 
$f < 0.15$ (i.e. for the length periods  $\geq$ 7 nucleotides, cf. last paragraph of the previous Section), 
the C (or G) dominated sequences 
exhibit a low coherence even in this domain. The last (low) coherence is a secondary effect to the the former one (see 
above).\\

  It should be also noted that the C/G coherency has a strong burst (the peak marked by an arrow in the Fig. 14) 
in a narrow vicinity of frequency $f\simeq 0.33$. This resonance peak comes from a very low coherency background and corresponds to the {\it codons}' period T=3 nucleotides (T=1/f). Let us recall that a codon is a sequence of three adjacent nucleotides constituting the genetic code (a specific amino acid residue in a polypeptide chain). 
Therefore, one can speculate that the two complimentary DNA strands can have relatively strong coherence 
related to the genetic code content in the case of the C/G containing codons, whereas the large-scale chaotic 
coherence related to the large-scale chaotic order in the A/T containing codons can suppress the 
genetic coherence (cf. Fig. 14). 
However, this speculation leads us beyond the pure physical frames of present paper.

\begin{figure} \vspace{-1cm}\centering
\epsfig{width=.45\textwidth,file=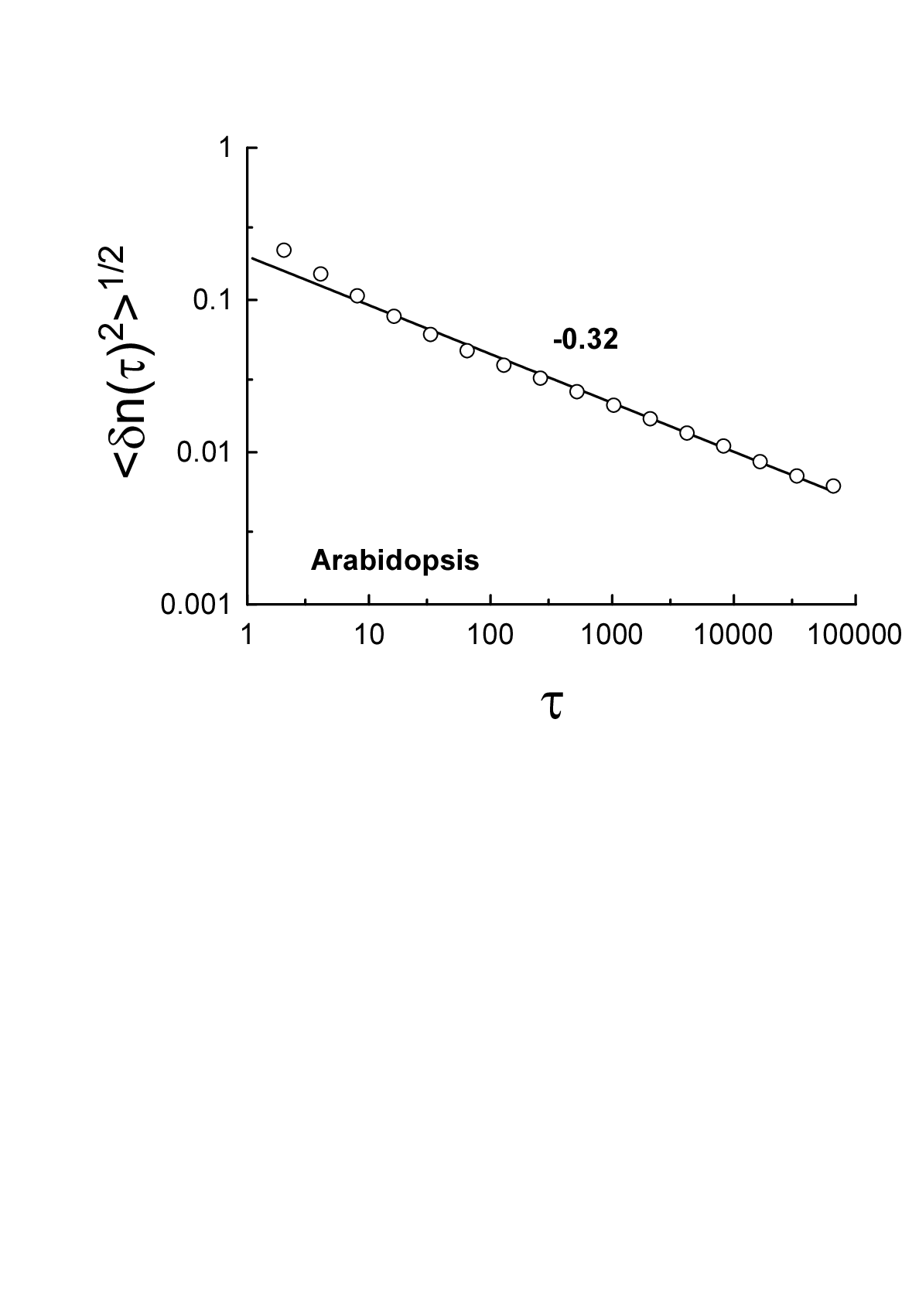} \vspace{-5cm}
\caption{The standard deviation for $\delta n (\tau)$ vs. $\tau$ for
the energy minima \{0,1\}-map used in section V (in log-log scales). 
The straight lines (the best fit) indicate the scaling law Eq. (2). }
\end{figure}

\section{Discussion}

For the Gaussian-like processes there is a very close relationship between their spectral and cluster-scaling properties \cite{b2}. If the system under consideration is non-Gaussian, then this relationship is broken. The
section V provides a good example of such situation. Indeed, figure 15 shows the standard deviation for $\delta n (\tau)$ vs. $\tau$ for the energy minima \{0,1\}-mapping used in the section V. The straight lines (the best fit) indicate the scaling law Eq. (2). Thus, for the considered non-Gaussian system a robust cluster-scaling (Fig. 15) 
can co-exist with the non-scaling spectrum (Fig. 10). Therefore, the long-range correlations (which correspond to the power-law scaling spectra) in the human genome \cite{voss}-\cite{pod} are not directly related to its cluster-scaling properties. These two types of scaling behavior are independent for the non-Gaussian systems that makes the 
cluster-scaling method an independent tool for studying these systems. Moreover, 
while the previously used for the spectral computations simple maps indicate stochastic behavior, the 
intrinsic map for the energy minima of stacking interactions indicates a chaotic (deterministic) interactions in 
an underground of the stochastic system on large scales. This large-scale chaotic order can be operational in the resolving the random gap problem mentioned in the Introduction and 
should affect the DNA-duplex dynamics creating, in particular, the large-scale coherence between the 
two strands of the DNA-duplex.

\end{document}